\documentclass[useAMS,usenatbib]{mn2e}

\usepackage{natbib}
\usepackage{graphicx}
\usepackage{graphics}
\usepackage[T1]{fontenc}
\usepackage{times}

\title[Centaurus\,A: filament nature from {\it XMM}]{Centaurus\,A: constraints on the nature of the giant lobe filaments from {\it XMM-Newton} observations}
\author[S. Wykes et al.]{Sarka Wykes,$^{1,2,3}$\thanks{E-mail: sw@extragalactic.info} Martin J. Hardcastle$^{4}$ and Judith H. Croston$^{5,6}$ 
\\
$^{1}$Department of Physics and Astronomy, University of Manitoba, Winnipeg, R3T 2N2, Canada\\
$^{2}$Department of Astrophysics/IMAPP, Radboud University Nijmegen, PO Box 9010, 6500 GL Nijmegen, The Netherlands\\
$^{3}$Anton Pannekoek Institute for Astronomy, University of Amsterdam, PO Box 94249, 1090 GE Amsterdam, The Netherlands\\
$^{4}$School of Physics, Astronomy and Mathematics, University of Hertfordshire, College Lane, Hatfield, Hertfordshire AL10 9AB, UK\\
$^{5}$School of Physics and Astronomy, University of Southampton, Highfield Campus, Southampton, SO17 1BJ, UK\\
$^{6}$Institute of Continuing Education, University of Cambridge, Madingley Hall, Madingley, CB23 8AQ, UK
}

\begin{document}

\date{Accepted 2015 September 21. Received 2015 September 2; in original form 2015 April 9}

\pagerange{\pageref{firstpage}--\pageref{lastpage}} \pubyear{2015}

\maketitle

\label{firstpage}

\begin{abstract}
We report on deep {\it XMM-Newton} observations of the {\it vertex} filament in the southern giant lobe of the Fanaroff-Riley class I radio galaxy Centaurus\,A. We find no X-ray excess from the filament region and place a $3\sigma$ upper limit on the $1$\,keV flux density of the filament of $9.6$\,nJy. This directly constrains the electron density and magnetic field strength in the filament. For the first time in an individual filament, we show that so long as the particle index $\ge2$, the excess in synchrotron emissivity cannot be produced purely by excess electrons: the filament magnetic field strength must be higher than in the giant lobes as a whole, and close to or above the equipartition value for the filament. The filaments are not significantly overpressured with respect to the surrounding lobe with a pressure provided by relativistic electrons. 
\end{abstract}

\begin{keywords}
acceleration of particles -- galaxies: individual (Centaurus\,A) -- galaxies: jets -- radiation mechanisms: non-thermal -- X-rays: galaxies.

\end{keywords}

\section{Introduction} \label{sect:introduction}

Centaurus\,A is a Fanaroff-Riley class I (FR\,I; \citealp{FAN74}) radio galaxy residing in the massive Ep galaxy NGC\,5128 \citep{HARR10b}. Its proximity of $3.8\pm0.1$\,Mpc \citep{HARR10a} makes it an attractive target for comprehensive physical studies with spatial resolution of $1$\,arcmin corresponding to $1.1$\,kpc. Centaurus\,A is a dynamically old object ($\sim560$\,Myr; \citealp{WYK13, EIL14, WYK14}), showing signs of multiple episodes of jet activity. The present-day jet represents a power of $P_{\rm j}\sim1\times10^{43}$\,erg\,s$^{-1}$ (e.g. \citealp{CRO09, WYK13}). Based on, independently, the energy required to expand the giant (i.e. outer) lobes and on the turbulent dissipation rate in them, \cite{WYK13} have found a time-averaged power of the pre-existing jet of $P_{\rm j}\sim1-5\times10^{43}$\,erg\,s$^{-1}$. A similar power range has recently been established by \cite{NEF15}.

Since the first radio isophotes of the giant lobes were reported by \cite{SHE58} and \cite{SHA58}, Centaurus\,A's large-scale structure has been extensively studied at radio frequencies, covering a wide range from $118$\,MHz to $41$\,GHz \citep{COO65, HAY83, JUN93, COM97, ALV00, ISR08, FEA09, HAR09, FEA11, MCK13, SUL13, STE13}. These studies revealed, for instance, lobe synchrotron ages of order $30$\,Myr \citep{HAR09}, relatively flat global spectral indices \citep{HAR09, MCK13}, the presence of substructure and some constraints on its origin \citep{FEA11}, and an upper limit on the thermal electron content of $n_{\rm e,th}<5\times10^{-5}$\,cm$^{-3}$ \citep{FEA09}. More recently, a contradictory, {\it tentative} measurement of $n_{\rm e,th}\sim1\times10^{-4}$\,cm$^{-3}$ was reported \citep{SUL13}, matching the estimated thermal particle content of the intragroup environment \citep{SUL13, EIL14} which the radio galaxy inhabits. Wykes et al.'s (2014) (W14) Giant Metrewave Radio Telescope (GMRT) observations at $\sim60\times40$\,arcsec spatial resolution, which we follow up on in this paper, have focused on the bright filamentary features in the southern giant lobe, the {\it vertex} (Largest Angular Scale, LAS, $\sim31$\,arcmin) and {\it vortex} (LAS $\sim58$\,arcmin). W14 established there that the pressure jumps over the filaments would translate to an acoustic Mach number of $\mathcal{M}\sim1.7$ ({\it vertex}) and $\mathcal{M}\sim1.6$ ({\it vortex}) based on minimum-energy assumptions, and to $\mathcal{M}\sim1.0$ for the more likely case of a stronger $B$-field in the filaments than in the global lobe plasma. They interpreted the origin of the {\it vertex} and {\it vortex} filaments in terms of weak shocks from transonic magnetohydrodynamical (MHD) turbulence or from the last vestiges of the activity of the pre-existing jet, or an interplay of both. In \cite{WYK13}, we proposed that the giant lobes became filled with baryonic material entrained in the current and pre-existing jet, and we predicted a thermal particle content of the lobes of $n\sim1\times10^{-8}$\,cm$^{-3}$.

Centaurus\,A is a strong case of dominant X-ray synchrotron in the jet on kpc scales (e.g. \citealp{GOO10}), a favoured X-ray emission mechanism on kpc scales in various other AGN jets (e.g. \citealp{HAR01, KRA05, WOR10, MEY15, GEN15}). Meagre information at X-ray frequencies exists on the giant lobes: {\it ASCA} (spatial resolution of $73$\,arcsec) pointing at an outer sector of the northern giant lobe probing the $0.5-2$\, and $2-10$\,keV continua resulted in a {\it tentative} detection of diffuse emission in the soft band with a flux $\sim8.5\times10^{-14}$\,erg\,cm$^{-2}$\,s$^{-1}\!/0.55$\,deg$^2$, best fitted with a thermal model with temperature $kT=0.6_{-0.8}^{+1.0}$\,keV, and a detection of a compact feature in the harder band modelled as a power law \citep{ISO01}; however, an association of the compact feature with a background source is probable (see also \citealp{STA13}). Stawarz et al.'s (2013) $\sim80$\,ks {\it Suzaku} observations (spatial resolution $\sim2$\,arcmin) in the energy range $0.5-10$\,keV provided {\it tentative} detection of a soft excess component with a flux $\sim6\times10^{-13}$\,erg\,cm$^{-2}$\,s$^{-1}\!/0.35$\,deg$^2$, most readily fitted by a thermal model with $kT\sim0.5$\,keV, corresponding to a thermal electron content of the lobes of $n_{\rm e,th}\sim1\times10^{-4}$\,cm$^{-3}$. However, the reported detection could also originate from the Galactic X-ray foreground \citep{STA13, WYK13}. Deep {\it INTEGRAL}-SPI observations (spatial resolution $2.5^{\circ}$) probing the $40$\,keV--$1$\,MeV continuum only resulted in a $3\sigma$ upper limit on photon flux of $<1.1\times10^{-3}$\,ph\,cm$^{-2}$\,s$^{-1}$ \citep{BEC11}. This limit is consistent with the detection of the giant lobes at gamma-rays with {\it Fermi}-LAT (spatial resolution $\sim1^{\circ}$) which culminated in measured photon fluxes of $(0.93\pm0.09)\times10^{-7}$\,ph\,cm$^{-2}$\,s$^{-1}$ (northern giant lobe) and $(1.43\pm0.15)\times10^{-7}$\,ph\,cm$^{-2}$\,s$^{-1}$ (southern giant lobe) after three years of all-sky monitoring (see \citealp{YAN12}; \citealp{ABD10} for the first ten-month result). The {\it Fermi}-LAT survey, in conjunction with high-frequency radio observations, has enabled estimation of the global magnetic field of the giant lobes of $B\sim0.9$\,$\mu$G.\footnote{However, see \cite{EIL14} for a discussion of $\mu$G or stronger magnetic fields; in her model, the radio emission likely originates from a different electron population than the gamma-ray emission.}

Mature FR\,I lobes are likely to be turbulent, and \cite{HAR09}, \cite{SUL09}, \cite{WYK13}, \cite{EIL14} and W14 have addressed in some detail MHD turbulence in Centaurus\,A's giant lobes. Mildly sub-Alfv\'enic MHD turbulence, as advocated for the lobes by \cite{WYK13}, permits the existence of long-lived filaments. Observational evidence for filamentary FR\,I-type lobes is in increasing supply, e.g., 3C\,310 \citep{BRE84}, Hercules\,A (\citealp{DRE84, GIZ03}; Cotton et al., in preparation), Fornax\,A \citep{FOM89}, M\,87 \citep{OWE00, FOR07}, NGC\,193 \citep{LAI11}, B2\,0755+37 \citep{LAI11}, M\,84 \citep{LAI11}. MHD turbulence, in general, implies filamentary structure in synchrotron emission (e.g. \citealp{EIL89, HAR13, WYK13}) and, crucially, computations of synchrotron radiation require an input magnetic field strength. While the global $B$-field strength of the giant lobes is probably well constrained by the {\it Fermi}-LAT survey (see above), the knowledge of its filling factor and the distribution of electrons in the lobes remain in an unsatisfactory state. The number density of electrons also  dictates the emissivity from the inverse-Compton upscattering of the cosmic microwave background (CMB) photons. 

If the radiating particles and magnetic field globally are at equipartition, they cannot dominate the internal energy of FR\,I lobes \citep{CRO03, CRO08, CRO14}. A likely scenario is additional material from entrainment, heated to temperatures substantially above that of the surrounding intragroup gas \citep{CRO08, WYK13, CRO14, KOL15}.

The main objective of the present paper is to investigate, from an X-ray perspective, the character of the {\it vertex} filament in Centaurus\,A's giant lobes. This assists us to constrain the distribution of internal energy within the giant lobes, and it provides a more complete picture of localised particle acceleration in the large-scale lobes. Here, {\it XMM-Newton's} sensitivity and spatial resolution of $6$\,arcsec are very suitable. Moreover, the LAS of the {\it vertex} is well matched to the instrument's field of view, and the filament is at a sufficiently large distance from the AGN core that the nuclear and jet X-ray emission does not compromise our aims. A single pointing in the direction of the {\it vertex} allows us to measure any excess inverse-Compton emission from the {\it vertex} with respect to {\it locally measured} background (i.e. without any contamination from large-scale thermal or non-thermal emission from the giant lobes. This places a limit on (or allows a measurement of) the electron density in the filament. 

The remainder of the paper is structured as follows. In Section\,\ref{sect:obs}, we describe the {\it XMM-Newton} observations and data reduction. Results are presented in Section\,\ref{sect:results}. In Section\,\ref{sect:interpret}, we use the data to provide constraints on the electron densities and pressure in the {\it vertex} filament and connote wider implications for the lobes. We conclude in Section\,\ref{sect:summary}.

Throughout the paper, we define the energy spectral index $\alpha$ in the sense $S_{\!\nu}\propto\nu^{-\alpha}$. The photon index is $\Gamma = \alpha + 1$ and the particle index $p = 2\alpha+1$. All coordinates are J\,2000.0.

\section{Observations and data analysis} \label{sect:obs}

The {\it vertex} filament was observed with {\it XMM-Newton} EPIC instruments on $14 - 15$ February 2014 (OBSID 0724620301) with the medium filter inserted. The MOS cameras were in the full-frame mode and the pn camera in the extended full-frame mode. After filtering for high particle background levels, these observations yielded $133\,601$ and $133\,572$\,s, respectively, for the MOS cameras, and $131\,939$\,s for the pn camera. Additionally, we recovered $3\,831$ (MOS1), $3\,802$ (MOS2) and $2\,169$\,s (pn) from heavily radiation-impacted observations on $7$ January 2014 (OBSID 0724620201). The ODF files were processed using {\sc sas} version 14.0.0. 

The MOS data were filtered to contain single to quadruple events (PATTERN $\le12$) and a count threshold of $0.2$\,counts\,s$^{-1}$ (RATE $<0.2$), whereas the pn data set was filtered to only include single and double events (PATTERN $\le$ 4) and allowing a count threshold of $0.35$\,counts\,s$^{-1}$ (RATE $<0.35$). Both data sets were energy-filtered to include events in the range $0.3 - 8.0$\,keV.

Images were created from the reduced data with the {\sc sas} task {\tt evselect} before applying the task {\tt evigweight} to correct for vignetting. We excluded, by eye, $47$ contamining point sources.

Filter-wheel closed data sets for particle background subtraction were processed and filtered in the same manner as the source data sets. The events lists for each EPIC camera were filtered using the same PATTERN filters as the data files, and weighted using {\tt evigweight}. Task {\tt attcalc} was utilised on the closed-filter files to recast the events list to the physical coordinates matching the real data. We then established a scaling factor for each background data set to account for differences in the normalization of the particle and instrumental background between the source and filter-wheel closed data sets. The scaling factors were computed by comparing the $10-12$\,keV count rates for the source and background data sets. The background data products were scaled by this factor before performing background subtraction.
\begin{figure*}
\includegraphics[width=0.7\linewidth]{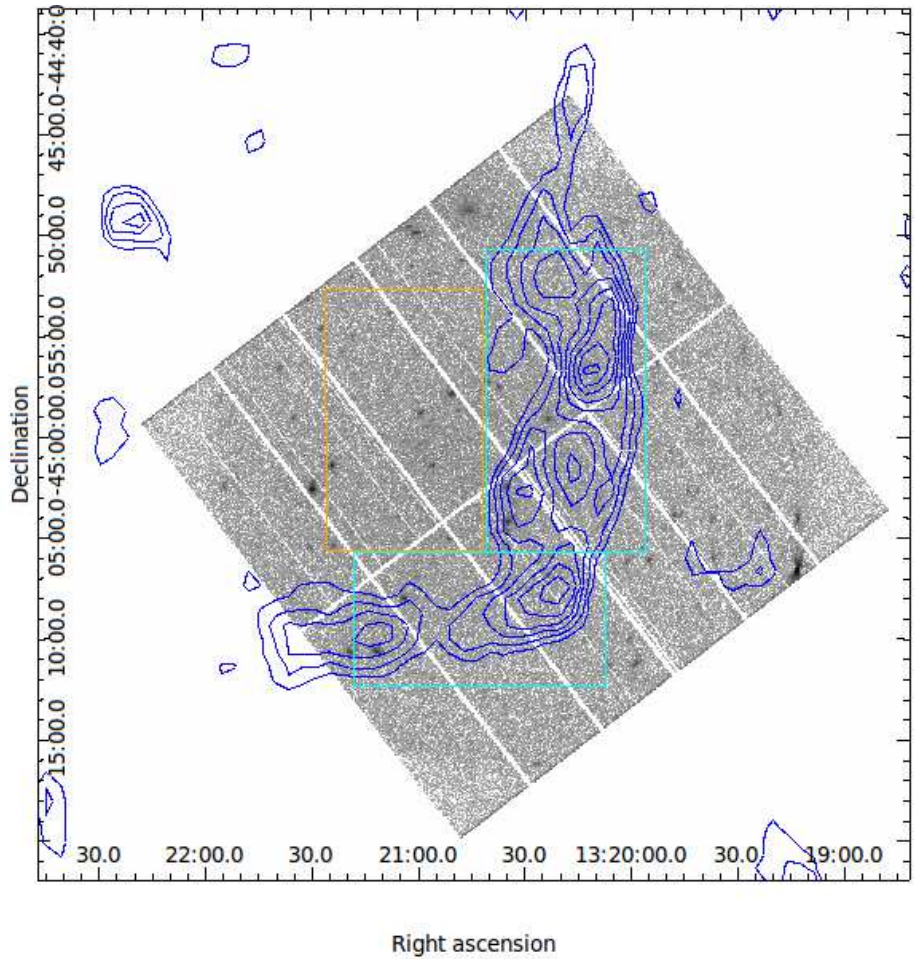}
\caption{$0.3-8.0$\,keV {\it XMM-Newton} image (pn, without vignetting correction) of the location of the {\it vertex} filament, with GMRT 325\,MHz radio contours for levels $6$, $8$, $10$, $12$, $14$, $16$, $18$ and $20$\,mJy\,beam$^{-1}$ overlaid. Also shown are the foreground and background regions described in the text. (A colour version of this figure is available in the online journal.)} \label{fig:fig1}
\end{figure*}

We then measured counts from the {\it vertex} region from the FITS files making use of {\sc ds{\footnotesize 9}}\footnote{\tt https://hea-www.harvard.edu/RD/ds9/site/Home.html} and the Funtools\footnote{\tt https://github.com/ericmandel/funtools} library, defining rectangular regions encompassing the filament segment positionally coincident with the detector areas. (Because of the instrument roll, i.e. a detector angular orientation not fully covering the filament extension on the sky, these are slightly smaller than the regions defined by W14.)

\section{Results} \label{sect:results}

Fig.\,\ref{fig:fig1} shows the X-ray emission from the  {\it vertex} region with GMRT radio contours at $325$\,MHz superposed. No X-ray excess from the filament regions is discernible.

We measure $2247\pm1210$\,counts in the combined pn observations in the $0.3-8.0$\,keV band in the approximate region of the {\it vertex} filament. Combining the livetimes and errors of the two observations, we obtain $(1.9\pm1.0)\times10^{-2}$\,counts\,s$^{-1}$. Since the errors are $1\sigma$, this is a $3\sigma$ upper limit of $3.1\times10^{-2}$\,counts\,s$^{-1}$ on the pn counts. Because of the missing MOS1 chips and the comparatively lower overall sensitivity of the MOS instruments, we exclude the MOS results from further discussion.

The upper limit on the count rate, with the medium filter inserted, a Galactic column density of $N_{\rm H}=6.6\times10^{20}$\,cm$^{-2}$ and an energy spectral index of $\alpha=0.5$, translates through WebPIMMS to a flux of $9.6$\,nJy.

\section{Interpretation} \label{sect:interpret}

In this section, we examine the limits placed by our observations on the physical conditions in the {\it vertex} filament, and the distribution of internal energy within the lobes.

\subsection{Filament electron density and $B$-field strength} \label{sect:density}

To evaluate the electron density and $B$-field strength in the {\it vertex} filament, we adopt a model with a fixed lower electron energy \citep{MYE85} and perform the synchrotron and inverse-Compton calculations numerically using the code of \cite{HAR98}, as in W14. The code assumes a fully tangled (isotropic) magnetic field and a
\begin{table*}
\begin{center}
\caption{Predicted inverse-Compton flux densities and total energy densities in the electrons and field as a function of the particle (electron) spectral index  $p$ and the length of the sampled area of the filament (with $l_1$ the projected length and $l_2$ twice the projected length).}
\label{tab:params}
\begin{tabular}{lccccc}
\hline
Model           & $p$  & \multicolumn{2}{c}{IC flux (nJy)} & \multicolumn{2}{c}{ $U_{\rm tot}$ (erg\,cm$^{-3}$)}\\
                &      & $l_1$ & $l_2$  & $l_1$ & $l_2$  \\
\hline
Equipartition   & $2.0$  & $1.43$ & $1.94$ & $5.17\times10^{-13}$ & $3.51\times10^{-13}$ \\
			    & $2.2$  & $1.69$ & $2.32$ & $7.54\times10^{-13}$ & $5.17\times10^{-13}$ \\
				& $2.4$  & $1.74$ & $2.42$ & $1.19\times10^{-12}$ & $8.26\times10^{-13}$ \\
				& $2.6$  & $1.65$ & $2.31$ & $1.93\times10^{-12}$ & $1.35\times10^{-12}$ \\
				& $2.8$  & $1.44$ & $2.04$ & $3.20\times10^{-12}$ & $2.26\times10^{-12}$ \\
\hline
$B=0.9$\,$\mu$G & $2.0$  & $7.64$  & $7.64$ & $1.41\times10^{-12}$ & $7.23\times10^{-13}$ \\
				& $2.2$  & $14.14$ & $14.14$& $3.19\times10^{-12}$ & $1.61\times10^{-12}$ \\
				& $2.4$  & $25.59$ & $25.59$& $8.77\times10^{-12}$ & $4.40\times10^{-12}$ \\
				& $2.6$  & $45.55$ & $45.55$& $2.66\times10^{-11}$ & $1.33\times10^{-11}$ \\
				& $2.8$  & $80.03$ & $80.03$& $8.87\times10^{-11}$ & $4.43\times10^{-11}$ \\
\hline
\end{tabular} \\
\begin{tabular}{l}
\scriptsize{Note. The radio data entries (GMRT $235$ and $325$\,MHz and ATCA $1.4$\,GHz flux
  densities) are taken from W14. The IC flux density  is at $2.4\times10^{17}$\,Hz. $E_{\rm e,min}=5\times10^6$\,eV, $E_{\rm e,max}=10^{11}$\,eV.}
\end{tabular} 
\end{center}
\end{table*}
uniform, energy-independent distribution of electron pitch angles. We adopt a distance of $3.8$\,Mpc to the {\it vertex}.\footnote{We assume that the giant lobes are in the plane of the sky and the {\it vertex} filament is in the lobe centre. The distance to the giant lobes is not well constrained; H\,{\small I} absorption against the south-west inner lobe and higher radio linear polarization of the northern inner lobe could hint at the northern giant lobe in front \citep{GOR90, JUN93}. The absence of a depolarization signal in the 1.4 and 5\,GHz radio continuum data at the position of the dwarf irregular galaxy ESO\,324-G024 (which is in projection on the northern giant lobe and at approximately the same distance as Centaurus\,A's core) also favours the northern giant lobe being in front \citep{JON15}. In turn, the southern giant lobe is both larger in projection and brighter (e.g. \citealp{ALV00, ABD10, FEA11}), which would suggest the southern giant lobe being closer if disregarding different physical conditions in the lobes.} 

The inverse-Compton emission from the filament is dominated by scattering of CMB photons, and so the results are essentially independent of the value of the low-energy cutoff, so long as it is below the energy $\gamma_{\rm e} \sim 1000$ for scattering of CMB photons into the X-ray (we adopt $\gamma_{\rm e,min} = 10$), and also of the spectral age, which affects the high-energy electron spectrum.\footnote{Appropriate values of the age to use range between $2$\,Myr -- the `turbulent age' of the {\it vertex}, i.e. its physical age if it is associated with turbulent eddies approximating the driving scale in the lobes (see W14), and at the higher end the value of $80$\,Myr -- taking into account the synchrotron age of $\sim30$\,Myr \citep{HAR09} and the upper limit on the radiative age of the lobes of $\sim80$\,Myr suggested by \cite{YAN12}. However, we verified that the inverse-Compton results were independent of choices in this range.} The extragalactic background light (EBL) inverse-Compton emission is about two orders of magnitude below that from the CMB for any reasonable EBL model (see \citealp{HAR09} and \citealp{ABD10} for details of models) for $p=2.0$, and $1.5$ orders of magnitude below for $p=2.8$, if we retain $\gamma_{\rm e,min} = 10$. For $\gamma_{\rm e,min} = 100$, the CMB inverse-Compton dominates over the EBL by even larger factors. Both \cite{HAR09} and Abdo et al. make the statement that the photon energy density in the giant lobes due to host galaxy starlight is much less than that of the EBL.

The key unknown parameter of the electron spectrum is the low-energy electron index $p$, corresponding to the injection index $\alpha_{\rm inj}$. \cite{MCK13} derived a global spectral index around the location of the {\it vertex} of $\alpha_{\rm 118\,MHz}^{\rm 1.4\,GHz}=0.63 \pm 0.01$, i.e. $p\sim2.2$, and in W14 we gave a best-fitting spectral index of the {\it vertex} from the GMRT data of $\alpha_{\rm 235\,MHz}^{\rm 1.4\,GHz}=0.81\pm0.10$, i.e. $p\sim2.6$. In light of these results, we adopt a range $p=2.0$ to $p=2.8$ for our modelling. Note that \cite{ABD10} have estimated $p=1.1$ to $1.6$ for the southern giant lobe based on a joint fit to the {\it Fermi}-LAT gamma-ray data and the radio data of \cite{HAR09}. However, these results are strongly dependent on the electron spectral model used; \cite{HAR09} showed that the same radio data can be fitted rather well with $p=2$ and a Jaffe-Perola aged spectrum \citep{JAF73} rather than the break and exponential cutoff in the electron energy assumed by Abdo et al. In addition, no error bars are quoted on the particle index by Abdo et al., and so it is unclear whether there genuinely is an inconsistency with $p = 2.0$ here. Their break is such that only the 408\,MHz data point constrains their `$s1$' index, and that has a comparatively large error bar; this potential large uncertainty on the value of $p$ derived from their fitting may explain the discrepancy between the low values of $p$ they obtain for the southern giant lobe and the much larger low-energy values ($p=2.1$) that they find for the northern giant lobe, which is difficult to understand in physical terms.

The estimated volume of the {\it vertex} segment that we consider, adopting cylindrical geometry with length $l=31$\,kpc and radius $r=3.2$\,kpc, is $V\sim2.9\times10^{67}$\,cm$^{3}$, but in order to estimate the effects of projection on the inverse-Compton calculation we also carry out the calculation on the assumption of a filament length (and so volume) double the measured value. Other input parameters are given in Table\,\ref{tab:params}.

The remaining parameter that affects the inverse-Compton emission is the magnetic field strength. We may consider three limiting cases of interest.
\begin{enumerate}
\item We can take the $B$-field strength to be that of the giant lobes, where the gamma-ray inverse-Compton emission detected by {\it Fermi}-LAT implies $B=0.9$\,$\mu$G. This would be appropriate if the filaments were purely electron (and possibly also non-radiating particle) overdensities in a constant (background) $B$-field.
\item We can assume that the filaments are in equipartition, i.e. the energy densities in radiating particles and $B$-field are the same. This condition approximately holds globally in the giant lobes, and also in the lobes and hotspots of some other (mostly FR\,II) radio galaxies, though the $B$-field energy density is usually a factor of a few to an order of magnitude below the electron energy density in these systems. Clearly, the higher volume emissivity of the filaments would imply a higher $B$-field strength than that in the giant lobes if the filaments were at or close to equipartition, and so, for a given synchrotron emissivity, would imply a lower inverse-Compton flux than for case (i).
\item We can assume that the filaments have the same {\it electron density} and spectrum as the giant lobes. In this circumstance, the filaments would be purely magnetic enhancements, and consequently the inverse-Compton emissivity would be no different from that of the giant lobes. This implies an even lower inverse-Compton flux than for case (ii), and in fact, we cannot even in principle detect X-ray emission from the filaments in this case, since they would have no contrast with the background provided by the lobes, to which we are not sensitive.
\end{enumerate}

Our inverse-Compton calculations therefore initially consider cases (i) and (ii) as limits. For case (ii), we calculate the equipartition $B$-field strength for each value of the electron index $p$ and filament volume that we consider: for case (i) we fix the $B$-field strength to $0.9\,\mu$G. The results are tabulated in Table\,\ref{tab:params}. (Note that under the $B=0.9$\,$\mu$G model, the values for inverse-Compton flux are equal for different filament length assumptions, since an increase in volume leads to a drop in the synchrotron emissivity, by the same factor, and to a decrease by the same factor of the value of $n_{\rm e,rel}$.)

Case (i), $B = 0.9\,\mu$G, is in conflict with our measured upper limit on the filament X-ray flux for essentially all values of $p$ (even the $p=2.0$ value, technically below the $3\sigma$ limit, is ruled out at better than $2\sigma$ confidence). We can thus conclusively rule out, so long as $p \ge 2.0$, a model in which the filaments are purely particle excesses. On the other hand, equipartitition models (case ii, with their lower electron densities) are all permitted by the X-ray limit.

The measured upper limit on inverse-Compton flux density sets a lower limit on the $B$-field strength, which depends on $p$, and we calculate this by varying the assumed fixed $B$-field to find the lowest value that satisfies the constraint $F\le9.6$\,nJy. The limits derived are shown in Table\,\ref{tab:blimits}, and are generally a factor $2-3$ below the equipartition value, interestingly close to the typical ratio $B \sim 0.7B_{\rm eq}$ observed globally in FR\,II radio lobes \citep{CRO05}. Our results thus suggest a filament $B$-field strength close to, or higher than, equipartition.

\begin{table}
\caption{Equipartition $B$-field strengths and inverse-Compton field strength limits as a function of $p$. The field strength is constrained by the $3\sigma$ flux limit to lie above $B_{\rm limit}$.}
\label{tab:blimits}
\begin{center}
\begin{tabular}{ccc}
\hline
$p$ & $B_{\rm eq}$ ($\mu$G) & $B_{\rm limit}$ ($\mu$G) \\
\hline
$2.0$ & $2.5$ & $1.0$ \\
$2.2$ & $3.1$ & $1.2$ \\
$2.4$ & $3.9$ & $1.5$ \\
$2.6$ & $4.9$ & $2.0$ \\
$2.8$ & $6.3$ & $2.5$ \\
\hline
\end{tabular}
\end{center}
\end{table}

\subsection{Filament pressure} \label{sect:pressure}

As we noted in W14, the particle and $B$-field content of the filament can have an effect on its dynamics. The pressures in the field and radiating particles can be derived from the energy densities in Table\,\ref{tab:params} by dividing by $3$. The (dominant) thermal pressure in the giant lobes was estimated by \cite{WYK13} to be $p_{\rm th}\sim1.5\times10^{-12}$\,dyn\,cm$^{-2}$ and by \cite{EIL14} to be $p_{\rm th}\sim3.2\times10^{-13}$\,dyn\,cm$^{-2}$; models for the filament where its non-thermal pressure significantly exceeds the thermal values (i.e. $U_{\rm tot} \ga 9 \times 10^{-12}$ erg cm$^{-3}$ for the higher thermal pressure value) are difficult to sustain as they would imply a strongly overpressured filament with a short lifetime (of order the sound-crossing time). We see (Table\,\ref{tab:params}) that the high-$p$ models with $B = 0.9$\,$\mu$G are in conflict with this constraint in addition to being ruled out by the inverse-Compton limits, while all equipartition models are permitted. $B$-field strengths lower than the equipartition value (Section\,\ref{sect:density}) would tend to increase the pressure but there is substantial room to do so without violating dynamical constraints at low-$p$ values.

\subsection{Other implications of the X-ray non-detection} \label{sect:energy}

The fact that we see no extended X-ray emission associated with the {\it vertex} places further constraints on the models of the origin of the large-scale filaments that we discussed in W14. In particular, if the filaments have been associated with thermal X-ray emission as seen from
the northern middle lobe by \cite{KRA09}, then models in which they were associated with the termination of a (possibly disconnected) large-scale jet would be favoured. However, no such emission is seen: nor is there any X-ray synchrotron emission such as is associated with the strong shocks currently being driven by the south-west inner lobe \citep{CRO09}. This strengthens the observational support for the favoured model of W14, namely that the filaments are weakly overpressured features driven by large-scale lobe turbulence. The radio spectrum and large-scale morphology of the southern giant lobe filaments has long been known to be different from that of the northern middle lobe \citep{JUN93,HAR09} and our X-ray non-detection adds to the evidence that their physical origin is quite different. We note that the idea of `middle lobes' presently connected to `large-scale jets' extending beyond the inner lobes in Centaurus\,A \citep{MOR99, SAN15}, always difficult to reconcile with the sharply bounded appearance of the inner lobes themselves, is disfavoured in the case of the northern middle lobe by the new $330$\,MHz observations of \cite{NEF15}.

\section{Summary} \label{sect:summary}

We have presented results from $\sim120$\,ks {\it XMM-Newton} observations of the {\it vertex} filament in the southern giant lobe of Centaurus\,A, aimed at determining the nature of the filament and the partition of internal energy within the lobes. The key results are as follows.

(1) No excess X-ray emission is detected from the filament. We find an upper limit ($3\sigma$) on the $1$\,keV X-ray flux density of the filament, assuming a power-law spectrum, of $9.6$\,nJy.

(2) We are able to rule out directly, for electron index $\ge2$, for the first time in an individual filament in any radio galaxy, a model in which the excess radio emissivity of the filament is caused purely by an excess of relativistic electrons in the presence of a constant background field. This supports less direct arguments from the overall appearance of lobes \citep{HAR05,GOO08}. Magnetic field variation is responsible for (at least some and possibly all of) the filamentary structures in radio galaxy lobes as a whole. This has important implications for age estimates, as spatially varying magnetic field would make spectral age estimates less reliable in determining the true physical age of a source.

(3) The X-ray constraints on magnetic field strength together with our modelling of our GMRT observations supports the model for the filaments presented by W14: they are likely to be at most mildly overpressured and to have their origin in transonic or subsonic turbulence in the giant lobes.

The remaining uncertainty in our modelling is the correct value of the particle index to use to extrapolate from the GMRT measurements -- high-resolution imaging of the giant lobes at the lowest possible terrestrial frequencies, tens of MHz, would be required to directly image electrons of the energies responsible for the X-ray inverse-Compton emission. The reference design for Phase\,$1$ of the Square Kilometer Array gives it the capability to constrain the low-energy spectral index in the filaments down to frequencies of $50$\,MHz. Much more sensitive X-ray observations would be required to detect the filaments if they are at equipartition or to show that their magnetic field strength exceeds the equipartition value: such observations should be possible using {\it ATHENA} towards the end of the next decade.

\section*{Acknowledgements} 
We thank Roger Clay and an anonymous referee for thoughtful comments. SW thanks the University of Hertfordshire for kind hospitality.

\bsp

\label{lastpage}

\end{document}